\documentclass[conference]{IEEEtran}
\usepackage{cite}
\usepackage{amsmath,amssymb,amsfonts}
\usepackage{algorithmic}
\usepackage{graphicx}
\usepackage{textcomp}
\usepackage{verbatim}
\usepackage{hyperref}

\usepackage{subfigure}
\usepackage{tabularx}

\usepackage[dvipsnames]{xcolor}
\usepackage{enumitem}

\newcolumntype{P}[1]{>{\centering\arraybackslash}p{#1}}
\newcolumntype{M}[1]{>{\centering\arraybackslash}m{#1}}
\newcolumntype{L}[1]{>{\raggedright\arraybackslash}m{#1}}

\def\BibTeX{{\rm B\kern-.05em{\sc i\kern-.025em b}\kern-.08em
    T\kern-.1667em\lower.7ex\hbox{E}\kern-.125emX}}
\begin{document}

\title{Trust in Software Engineering: \\ A Case Study on GitHub Pull Requests}
\title{Interpersonal Trust in OSS: Exploring Dimensions of Trust in GitHub Pull Requests}


\author{Double Blind Submission}

\author{\IEEEauthorblockN{Amirali Sajadi}
\IEEEauthorblockA{
\textit{Drexel University}\\
Philadelphia, Pennsylvania, USA \\
amirali.sajadi@drexel.edu}
\and
\IEEEauthorblockN{Kostadin Damevski}
\IEEEauthorblockA{
\textit{Virginia Commonwealth University}\\
Richmond, Virginia, USA \\
kdamevski@vcu.edu}
\and
\IEEEauthorblockN{Preetha Chatterjee}
\IEEEauthorblockA{
\textit{Drexel University}\\
Philadelphia, Pennsylvania, USA \\
preetha.chatterjee@drexel.edu}
}

\maketitle

\begin{abstract}

Interpersonal trust plays a crucial role in facilitating collaborative tasks, such as software development. 
While previous research recognizes the significance of trust in an organizational setting, there is a lack of understanding in how trust is exhibited in OSS distributed teams, where there is an absence of direct, in-person communications. 
To foster trust and collaboration in OSS teams, we need to understand what trust is and how it is exhibited in written developer communications (e.g., pull requests, chats). 
In this paper, we first investigate  various dimensions of trust to identify the ways trusting behavior can be observed in OSS. 
Next, we sample a set of 100 GitHub pull requests from Apache Software Foundation (ASF) projects, to analyze and demonstrate how each dimension of trust can be exhibited. Our findings  provide preliminary insights into cues that might be helpful to automatically assess team dynamics and establish interpersonal trust in OSS teams, leading to successful and sustainable OSS.

\end{abstract}

\begin{IEEEkeywords}
trust, open source software, pull requests
\end{IEEEkeywords}

\section{Introduction}
Although Open Source Software (OSS) has revolutionized the software development industry, stakeholders still face challenges in increasing the likelihood of new  projects' success and maintaining activity in existing projects~\cite{9402044, yin2021sustainability}. 
Studies suggest the inexorable importance of the health of the project participants towards the sustainability of OSS projects ~\cite{zanetti2013riseandfall}. For instance, McDonald and Goggins found that project success in OSS is attributed more to contributor growth and community involvement than to source code metrics~\cite{mcdonald2013performance}. Issues stemming from person/team dynamics, e.g., conflict between developers, is identified as one of the key reasons for failures in OSS projects~\cite{coelho2017why}. Interpersonal trust could facilitate collaboration, and lower the conflicts among team members~\cite{calefato2017preliminary, calefato2016affective}. Additionally, given the evidence suggesting that the newcomers' decision to leave OSS projects is influenced by their interactions with other team members~\cite{steinmacher2013newcomers}, there is an opportunity in studying how building trust in early stages can help with the retention of newcomers in OSS.

Having been approached from a broad range of disciplines, human trust has been studied in various contexts \cite{rusman-foster-trust}; but, the role of interpersonal trust in the context of software engineering has yet to be sufficiently examined and explored. 
Trust has been shown to influence the ability of a team to have better cohesion and more cooperation \cite{acedo2014personal}, to affect the developers in decision making \cite{orsila2009trust} and to retain new contributors in projects \cite{lane2004interpretative}. Furthermore, trust among team members has been used to determine the overall efficiency and the success of OSS development teams, while reduced levels of trust have been associated with a lack of unity and less collaboration \cite{calefato2017preliminary}.

Researchers have investigated trust, and how different types of trust (e.g., cognitive trust, dependency trust) are exhibited in an organizational setting~\cite{wierzbicki2010trust, rusman-foster-trust, khan-role-of-trust, calefato2016affective, network-centric}. Trust in the context of team management has been defined by multiple studies, some of which are listed in Table \ref{tab:trust_definitions}. However, there is no consensus on a universal definition of trust \cite{mystery, rousseau1998not}. 
To foster trust among developers and better manage teams, we first require a deeper understanding of what trust is and how it is manifested in the software development environments.



    

\begin{table}
    \centering
    \footnotesize
    \caption{Trust, as defined in the context of team management.}
    \vspace{-.2cm}
    \begin{tabular}{|L{0.46\textwidth}|}
    \hline
         ``
         the overall willingness of virtual team members to rely on one another that results from the aggregate of potential trust dimensions that are achieved through socio-emotional and task process and supported by technology capabilities." \textcolor{MidnightBlue}{Mitchell \& Zigurs \cite{mystery}} \\
    \hline
         ``the willingness of a party to be vulnerable to the actions of another party based on the expectation that the other party will perform a particular action important to the trustor, irrespective of the ability to monitor or control that other party” \textcolor{MidnightBlue}{Mayer et al. \cite{mayer1995integrative}} \\
    \hline
         ``a positive psychological state (cognitive and emotional) of a trustor (person who can trust/distrust) towards a trustee (person who can be trusted/distrusted) comprising of trustor’s positive expectations of the intentions and future behaviour of the trustee, leading to a willingness to display trusting behaviour in a specific context." \textcolor{MidnightBlue}{Rusman et al. \cite{rusman-foster-trust}} \\
    \hline
         ``an actor’s expectation of the other actors’ capability, goodwill and self-reference visible in mutually beneficial behaviour enabling cooperation under risk" \textcolor{MidnightBlue}{Henttonen \& Blomqvist \cite{henttonen2005managing}} \\
    \hline
    \end{tabular}
    \label{tab:trust_definitions}
    \vspace{-0.5cm}
\end{table}

In this paper, we address that gap by exploring how interpersonal trust is exhibited in OSS. First, we leverage theories from psychology and organizational behavior to investigate dimensions of trust. Each of the trust dimensions we examined is exhibited in specific types of interpersonal interactions, while the overall trust between a pair of individuals is the accumulation of all these  dimensions \cite{mystery}. With the goal of representing instances of trust in OSS, we investigate pull requests on GitHub and create a mapping between the trust dimensions and activities in pull requests (Section \ref{sec:trust}). Next, we 
provide qualitative and quantitative support for the dimensions of trust we observed in a set of 100 GitHub pull requests (Section \ref{study}). 
Our results provide preliminary indications regarding the dimensions of trust that are exhibited in OSS, specifically in pull requests, and also provide insights into what cues might help in automatically extracting that information. The overarching goal of this study is to bring to attention the importance of interpersonal trust in OSS sustainability.

\begin{table*}[t]
    \centering
    \footnotesize
    \caption{Trust Dimensions in GitHub Pull Requests.}
    \vspace{-.2cm}
    \begin{tabular}{|m{.425\textwidth}|m{.525\textwidth}|}
    \hline
         \textbf{{\em This dimension of trust develops from...}} & \textbf{{\em For Pull Requests, this dimension of trust can be observed in...}} \\
    \hline
    {\em \textcolor{MidnightBlue}{Action-based trust}} -- ``a process based on or promoted by fast and frequent feedback with minimal delay."
    \cite{alexander2002teamwork, mystery}.& The frequency and quality of comments and reviews of contributed pull requests can form a basis for action-based trust between the contributor and the reviewer.\\
    \hline
    {\em \textcolor{MidnightBlue}{Commitment trust}} -- ``a process based on contractual agreements (formal or psychological) between parties who have an expectation of mutual benefit derived from cooperative relations." \cite{mystery} &  Reviewing pull requests when requested or   addressing a reviewer's comment indicates the existence of commitment trust between the contributor and reviewer.\\
    \hline
    {\em \textcolor{MidnightBlue}{Competence trust}} -- ``a process based on perceptions of another’s competence to carry out the tasks that need to be performed and is based on an attitude of respect for the abilities of the trustee to complete their share of the job at hand." \cite{mystery} & Developers' past actions in the repository, their overall activity, and their status on GitHub (e.g., previous pull requests, role in the project, level of activity, number of followers, etc.) influence the trustor's perception of the trustee's competence. 
    \\
    \hline
    {\em \textcolor{MidnightBlue}{Institutional trust}} -- ``a process guided by the norms and rules of institutions (such as organizations) and based on formal institutional arrangements such as contracts, sanctions, or legal procedures." \cite{mystery} & Developers associated with the same organizations (or working for the same company) are more likely to trust each other on the basis of their affiliations with these organizations. \\
    \hline
    {\em \textcolor{MidnightBlue}{Personality-based trust}} -- ``a developmental process that occurs during infancy when a person seeks help from caretakers and that results in a general propensity to trust others." \cite{mystery} & Project developers/maintainers who tend to accept a high number of pull requests from various users have a higher propensity to trust people and their work. On the other hand, those who tend to reject more pull requests may have a lower propensity to trust others.\\
    \hline
    {\em \textcolor{MidnightBlue}{Transferred trust}} -- ``a process that may occur when the trustor knows and trusts a person or the institution that recommends the trustee." \cite{alexander2002teamwork, mystery} & In some cases, well-established members of the project can recommend the work of new and less well-known contributors. This leads to other project participants who trust them to now trust the new contributor.
    \\
    \hline
    \end{tabular}
    \label{tab:trust_dimensions}
    \vspace{-0.4cm}
\end{table*}

\section{Trust dimensions}\label{sec:trust}

We select from the trust dimensions aggregated in the survey by Mitchell and Zigurs \cite{mystery} with the goal of representing instances of trust in OSS. 
Using the definition of each trust dimension, we selected the dimensions of trust that are not conflicting or overlapping with one another, and are applicable to OSS projects on GitHub. 
The trust dimensions we selected are as follows:


\noindent
\textbf{Action-based trust}: 
    One of the characteristics of high-trust teams is frequent and detailed communication~\cite{anybody-out-there}. While feedback in teams with lower levels of trust is usually limited to a few words (e.g., `ok', `looks good'), teams with high-trust 
    provide detailed feedback to each other's contributions.

\noindent
\textbf{Commitment trust}:
    Commitment trust is tied to the extent to which members of a team perceive themselves as a part of a team \cite{newell2007exploring}.
    The more engaged and trusted the developers are, the more committed they are to the project.

\noindent
\textbf{Competence trust}:
    The competence and the overall abilities of an individual to perform certain tasks within a specific context is closely related to how trustworthy they may seem to others \cite{rusman-foster-trust, kuo2009exploratory}. The trust inspired by one's perceived competence in a virtual team often depends on the data that can be publicly observed for each user. 

\noindent    
\textbf{Institutional trust}: People who work in the same environment led by the same set of rules are more likely to trust one another \cite{sarker2003virtual}. Belonging to certain known group or organization may lend a developer a degree of institutional trust.

\noindent    
\textbf{Personality-based trust}: An individual's general tendency to trust others, which is independent of external factors (e.g., the context), defines their personality-based trust. Those with a higher propensity to trust are more likely to trust others' ideas or contributions. 

\noindent    
\textbf{Transferred Trust}: Most people view individuals recommended by someone they trust in a positive light. This positive attitude toward the recommended person is called transferred trust. With transferred trust, the individual extends her trust from the person she already knows to the recommended party. 

\section{Methodology}
In order to understand trust in OSS projects and to perform our analysis, we focus on GitHub pull requests as they are a natural place to look for trust, specifically for the dimensions of trust we selected (in Section \ref{sec:trust}). Having considered the data available on GitHub pull requests, we examined the features that support pull requests and the type of developer interactions they tend to elicit. Doing so allowed us to look for clear signals of trust between the participants (i.e., contributors and reviewers).
More specifically, we considered the following types of information to understand how the trust dimensions are exhibited:
\begin{itemize}[leftmargin=*]
    \item {\em Pull request comments:} pull request comments are likely to contain valuable insight into the relationships of developers interacting with one another. 
    \item {\em Pull request metadata:} For each pull request, we examined the following information: pull request creator, closer, the state of the pull request (accepted or rejected), the state of the reviews (approved, commented, changes requested), the labels, and the type of contribution (code or documentation).
    \item {\em User profile data:} A GitHub user's publicly available profile data can influence their social interactions and the trust that develops between them and other project participants. We examined the following data:
    1) the number of followers, 2) the names of GitHub organizations the user belongs to, 3) previous actions in the repository of interest. 
\end{itemize}

To identify trust in the pull requests in our dataset, we adopted a thematic approach and followed the steps recommended by Nowell et al.\cite{nowell2017thematic}. 
The process of analyzing the data was done iteratively and reflectively, alternating between different phases of analysis such as memoing, identifying themes, and refining, as recommended in qualitative analysis~\cite{qual}.  
First, in order to better understand the interactions of the involved developers and the circumstances under which they have communicated, all the pull request comments and reviews were carefully read while keeping in mind the actions taken by the users to address the pull requests. Next, we used all the available data, including text, the background of the users (as available on their GitHub profiles), and the collected statistical data, to generate our initial set of mappings of trust dimensions with GitHub pull requests. These mappings indicated the presence or lack of a trust dimension in a data instance. 
We also made notes on the possible reasons for the pull request rejections as well as any relevant fact that could become useful in the following steps. We repeated the process multiple times, adding to our notes and observations, and reevaluating the mappings in each iteration.

In, Table \ref{tab:trust_dimensions} we show the result of this process, i.e., our mapping between the trust dimensions and the available data surrounding GitHub's pull requests.

\section{Preliminary Study}
In this section, we discuss our observations for the dimensions of trust we examined in a sample of GitHub pull requests. 
\label{study}

\noindent
\textbf{Data Selection.}
We selected four active repositories ({\tt superset}, {\tt beam}, {\tt airflow}, and {\tt dubbo}) from Yin et al.'s dataset of 269 Apache Software Foundation (ASF) projects \cite{yinapache}. Each repository in our dataset contains at least 5900 stars, 400 contributors, and 5000 pull requests. 
For the purpose of this study, we randomly sampled a set of 100 pull requests, 25 from each repository. 
All four repositories we examined contain predominantly accepted pull requests and few rejected pull requests. 
Therefore, to investigate the differences in the levels of trust exhibited in accepted vs. rejected pull requests, we sampled 75\% accepted and 25\% rejected pull requests. 

\noindent
\textbf{Observations.}
We examined the pull requests in our sample for instances of trust (expressed through the dimensions in Table \ref{tab:trust_dimensions}).
In Figure \ref{fig:trust_examples}, we visualize two example pull requests where: (a) one is exhibiting high levels of trust, (b) one is exhibiting low levels of trust. We report our observations on how each trust dimension manifests in pull requests below:

\begin{figure*}
    \centering
    \subfigure[ A pull request interaction with high levels of trust.]{\includegraphics[width=0.49\textwidth, height=5.8cm]{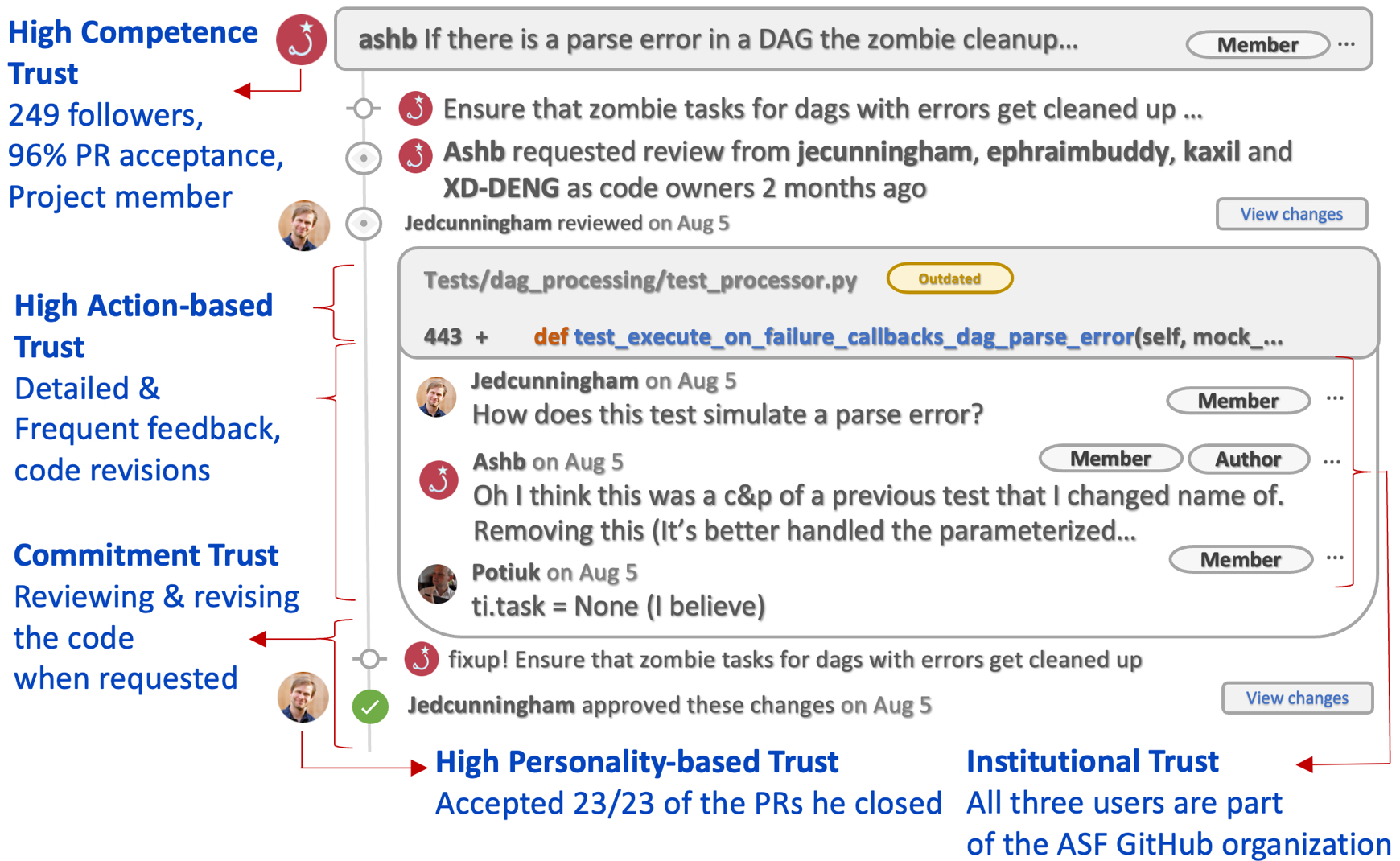}\label{fig:high-trust}}
    \subfigure[A pull request interaction with low levels of trust.]{\includegraphics[width=0.49\textwidth, height=5.8cm]{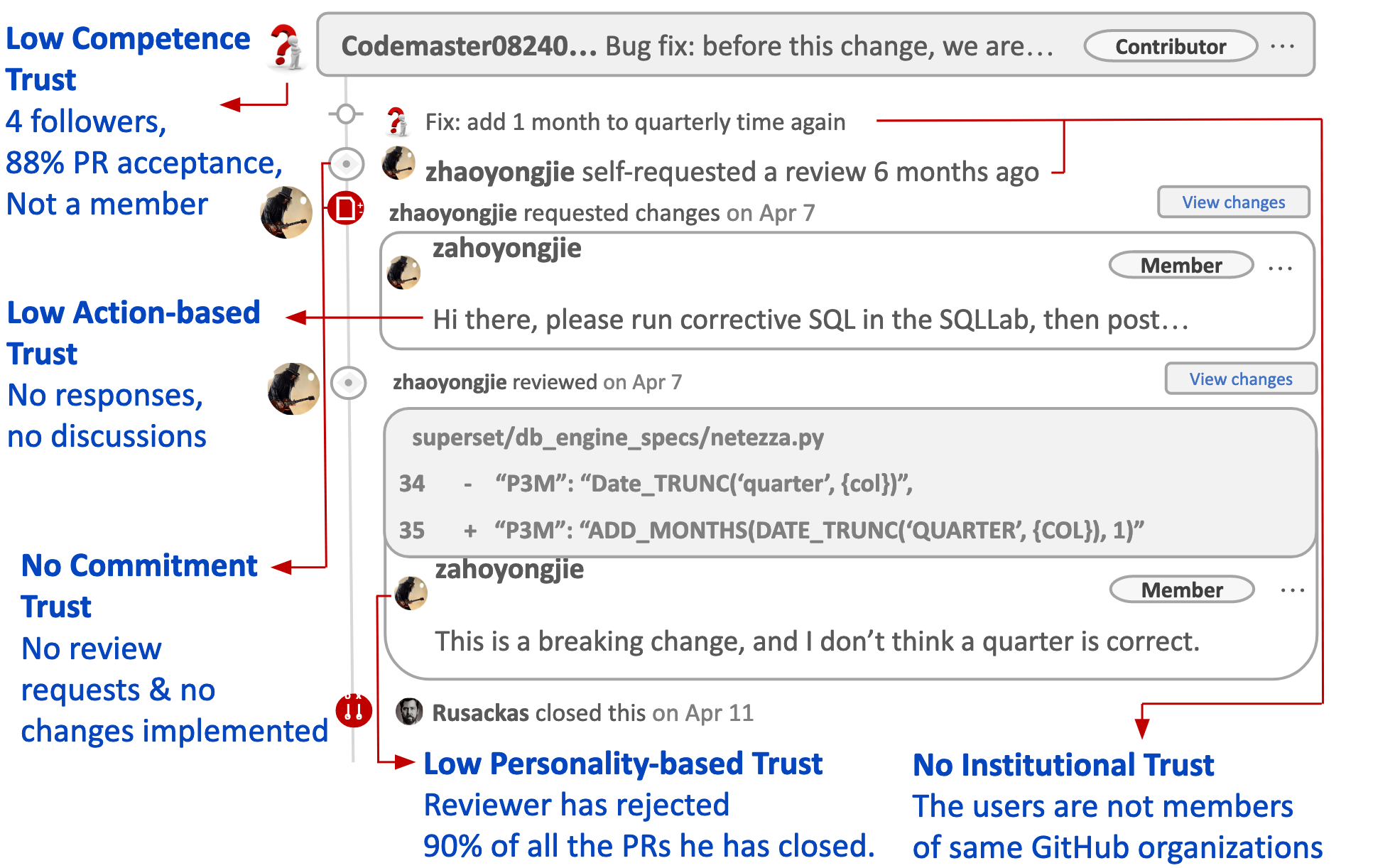}\label{fig:low-trust}}
    \caption{Examples of Trust exhibited in GitHub pull requests.}
    \label{fig:trust_examples}
    \vspace{-0.4cm}
\end{figure*}

\subsubsection{Action-based trust}
We found numerous instances of developer interactions in pull requests that resemble action-based trust. We observe that, compared to rejected pull requests, accepted pull requests generate more social interactions (e.g., comments) in a shorter period of time. Additionally, the creators of accepted pull requests are often more responsive to other users' questions. They actively try to explain or improve the quality of the contribution by asking for feedback and reviews from other developers. It is therefore more common to see multiple new commits and revisions in accepted pull requests, compared to the rejected ones. 
In our dataset, the average comment frequency (the number of comments in a day) is 4 for the accepted pull requests, and $\sim$1.25 for rejected pull requests. 75 out of 100 pull requests involved additional commits that were made as a result of developers' interactions. 66 of the accepted pull requests contributed to this number, while only 9 rejected pull requests contained additional commits made as a result of further communication.

Overall, the number and frequency of the comments and the  conversations leading to new commits have been used as indicators of action-based trust in the pull requests. For instance, the pull request in Figure \ref{fig:high-trust} has generated a high frequency of comments and created discussions that have resulted in revisions to the code. The pull request in Figure \ref{fig:low-trust}, however, has not received as much attention in the comments section, nor has it received any reviews that would aid the contribution to be further made suitable 
for merging.

\subsubsection{Commitment trust}

Given the regularity and the emphasis on the reviews in pull requests, we observe several instances of users requesting and making pull request reviews in our dataset.  It is uncommon for a requested user to make reviews after another requested user has already reviewed the work. This behavior could be attributed to the high volume of requests and the fact that the pull requests may not need multiple review approvals to get merged. When developers ask for reviews from multiple users, most often at least one of the requested users provides a review. 35 out of 100 pull requests in our dataset (32 accepted and 3 rejected pull requests) had at least one response to the requested reviews. 

The pull request creator in Figure \ref{fig:high-trust} requests four developers for reviews and receives a detailed review from one. On the other hand, in Figure \ref{fig:low-trust} 
the pull requester does not request a review or respond to the other user. 

\subsubsection{Competence trust}
Among the different profile attributes that we considered (number of followers, number of pull requests made and their acceptance/rejection rate in the last 1000 pull requests, write/read permission in the repository), the perceived trustworthiness of a  user is found to be closely related to their previous actions.  
Out of the 24 rejected pull requests, 9 were made by users who had not previously made any accepted pull requests in the project. Additionally, 4 of these 9 users were making their first pull requests in the project, while the other 5 had only made 1 rejected pull request prior to that point. 

In Figure \ref{fig:high-trust}, \texttt{ashb}, the pull requester, is a member of the project with write access in the repository, has a noticeable number of followers, and a high rate of acceptance for his pull requests; all qualities that would indicate competence trust. The pull requester in Figure \ref{fig:low-trust} has also made a considerable number of accepted pull request in the project, but compared to the requester in Figure \ref{fig:high-trust}, he has a substantially less number of followers and is not a project member.

\subsubsection{Institutional trust}
Inspecting the overlap between the developers' GitHub organizations, we observed that pull requests involving users with same affiliations 
has often led to successful contributions to the project (e.g., contributions that were not rejected on the spot). 
It should be noted that although GitHub organizations can help us detect a user's associations, not all organizational affiliations are available on GitHub. 
Additionally, an user can trust others based on their affiliations with certain organizations without being part of that organization. 
Out of the 25 rejected pull requests in our dataset, we did not observe any commenters or reviewers sharing the same organizations as the pull requesters. For accepted pull requests, we observed institutional trust in up to 11 pull requests out of the 19 accepted pull requests in each repository using the GitHub organizations.

The pull requester and both of the reviewers in Figure \ref{fig:high-trust} are parts of the ``The Apache Software Foundation" GitHub organization. No such cues for institutional trust is observed in the pull request in Figure \ref{fig:low-trust}.

\subsubsection{Personality-based trust}
 We can estimate the personality-based trust of a developer using the number of pull requests rejected or accepted. Surprisingly, most developers who closed pull requests have almost always accepted them. Specifically, 79 out of 100 pull requests had a closer or at least one reviewer who had accepted all the pull requests she had closed. For instance, we found that a specific user closed 272 pull requests, all of which were accepted. This might be due to the fact that each of the projects we examined are managed by small groups of active developers who contribute frequently and have most of their pull requests accepted. This pull request acceptance behavior and team dynamics is common across many OSS repositories~\cite{steinmacher2013newcomers}. More interestingly, we found that there are a set of users who predominantly reject the pull requests they close. Even though these users interact with other developers in different contexts (e.g., merged pull requests), the pull requests they close are almost always rejected.

The reviewers \texttt{jedcunningham} and \texttt{potuik} in Figure \ref{fig:high-trust}, almost always accept the pull requests they close, which we interpret as a sign of a high propensity to trust other developers' work. Conversely, \texttt{zhaoyongjie}, the reviewer in Figure \ref{fig:low-trust}, has only accepted approximately 10\% of all the pull requests they have closed (10 out of 99).

\subsubsection{Transferred trust}
We observed only one instance of transferred trust in our dataset. A significant contribution by a newcomer caused one of the project members to seek another team member's opinion on the work\footnote{https://github.com/apache/airflow/pull/25324}. The team member responded by introducing the pull requester as a new member of their team and assured the other party of the work's quality; \textit{``Syed is a new member of our team, we already reviewed his work :) Thanks for reaching out though"}. Overall, we require more data to better understand and observe this dimension of trust and the way it is exhibited on GitHub.

\section{Related Work}
\label{related_work}
From human trust in automation systems to designing better trust management \cite{wierzbicki2013improving} and e-commerce systems that find the best matches between customers and sellers based on trust \cite{hoff2015trust}, prior research has approached the concept of trust from various angles and in different capacities. Trust in software, systems and the ties between trust, privacy, and security are some of the  ways trust has been studied in a software-related context \cite{lavazza2010predicting, ozkaya2022understanding}. Some studies have examined trust as an interpersonal concept, trying to shed light on how trust is developed and how it affects the developers in  software engineering and open source teams \cite{al2011understanding, paul2012time, anybody-out-there, rico2009joint, rusman2012can, corbitt2004comparison}. Some of these studies use questionnaires or interviews to identify and understand trust in an organization and teams, while others have proposed methods to estimate the levels of trust in indirect ways. Wang \& Redmiles, for instance, collect data about trust through interviews to gain insight into the development of trust among software developers \cite{wang2016cheap}, while da Cruz et al. \cite{arsenal-gsd} puts forth a framework, meant to automatically detect trust among OSS project members. Venigalla \& Chimalakonda use an emotion analysis tool, and Sapkota et al. create a framework to automatically identify trust in Github \cite{venigalla2021understanding, network-centric}. Another study tries to evaluate developers' propensity to trust others by making use of IBM Watson's Tone Analyzer  and the \textit{Big-Five personality} model \cite{big-five}, a famous taxonomy of human personality traits, to find correlations between one's tendency to trust others and the chances of her accepting or rejecting pull requests. Likewise, Iyer et al. draws conclusions about propensity to trust in developers and the ways in which it can affect the chances of a pull request to get accepted or rejected \cite{iyer2019effects}. Lastly, Calefto \& Lanubile use sentiment analysis to build upon Wang \& Redmiles's results and estimate trust among developers \cite{calefato2016affective}. 
Apart from trust, researchers have analyzed and designed techniques for the detection of various emotions such as anger, stress, toxicity in communications related to software development~\cite{imran2022, Chatterjee21, Chatterjee2019, miller2022did, VADSE}. 
Compared to other work, we have investigated trust in OSS at a different granularity, by leveraging dimensions of trust based on theories in organizational behavior and psychology.

\section{Threats to Validity}
\textbf{Construct Validity.} The complex nature of interpersonal trust does not allow us to precisely determine the level of trust or lack thereof among individuals based on a limited number of written interactions. This limitation poses an unavoidable threat to the construct validity of the our study. Building upon prior theoretical work, we have attempted to address the inherent complications of identifying interpersonal trust in GitHub’s OSS projects. We use the same definitions of trust dimensions that have been proposed in the existing body of literature to ensure that the studied measures are, in fact, indicative of the answers we seek.

\textbf{Internal Validity.} The human interactions among those involved with any OSS project may not be limited to  pull requests. Even within the GitHub platform itself, there are other ways in which developers typically interact with one another. Furthermore, the developers can contact each another through different means and platforms, e.g., email or Slack. It is even possible that some developers know each other personally and have interactions outside the development environment. These factors can all affect the conclusions from our preliminary study. 
To mitigate this threat, we have limited our subjects of study to ASF projects. This measure limited the communication channels outside GitHub that the developers can use to contact each other as ASF projects follow specific communication guidelines. The ASF project members use mailing lists and GitHub features such as Discussions and Issues to communicate. Examining all these other communications is beyond the scope of this study and remains to be further explored in future research.

\textbf{External Validity.} In presenting the ways in which trust is manifested on GitHub, we have used a dataset of 100 randomly selected pull requests from four active repositories, which is a threat to generalizability. To mitigate this threat, we selected the subjects of our study from 4 popular OSS projects. Scaling to larger and more diverse datasets might lead to different study observations. We hope to assess the generalizability of our results by analyzing a larger dataset as a part of our future work. 

\section{Conclusion and Future Work}
The success of an OSS team effort is heavily influenced by the interpersonal dynamics of the developers. In this work, we sought to elucidate the crucial role of interpersonal trust among the OSS developers and to identify and illustrate this trust within the development context through our mapping of trust dimension to the GitHub interaction (Table \ref{tab:trust_dimensions}).
 The end goal of our research is to design an automatic system to detect and analyze trust in OSS teams. 
 Such automation would allow the monitoring of a team's level of trust and timely interventions for project managers. Furthermore, the measurements of trust can help us uncover ties between trust and the overall performance of a team and thus aid the tracking  the progress of the project throughout the software development lifecycle. 
 
Some of the cues we have used to determine trust 
can be quantitatively measured, leading to automatically computable measures (e.g., measuring \textit{Competence trust} from developer profile). However, other signals of trust, such as the textual data analyzed to detect \textit{Transferred trust} are not easily computable and might require more sophisticated natural language analyses. 
Of course, a larger study is needed to gain a deeper understanding and overcome these challenges. 
 As a part of a future study, we also plan to expand the dataset to include the ASF mailing list, which offers a rich source of data for developer communications beyond pull requests. The email addresses of the members can also enable us to detect members of the same team and better identify \textit{Institutional trust}. Our replication package is publicly available  \href{https://figshare.com/articles/dataset/trust_paper_-_pull_requests_dataset_csv/21330714}{here}.


\section{Acknowledgements}
This work was supported in part by Drexel University Summer Research Award.

\bibliographystyle{plain}
\bibliography{citation}

\end{document}